\documentclass[conference]{IEEEtran}
\IEEEoverridecommandlockouts
\usepackage{cite}
\usepackage{amsmath,amssymb,amsfonts}
\usepackage{algorithmic}
\usepackage{graphicx}
\usepackage{textcomp}
\usepackage{xcolor}

\usepackage{comment}
\usepackage{bm}
\usepackage{subfigure}

\usepackage{fancyhdr}
\def\BibTeX{{\rm B\kern-.05em{\sc i\kern-.025em b}\kern-.08em
    T\kern-.1667em\lower.7ex\hbox{E}\kern-.125emX}}
\begin{document}

\title{Non-contact Cold Thermal Display by Controlling Low-temperature Air Flow Generated with Vortex Tube
\thanks{*This work was supported by JSPS KAKENHI Grant Numbers JP17H01780.}
\thanks{$^{1}$Jiayi Xu and Osamu Oshiro are with the Graduate School of Engineering Science, Osaka University, Osaka, Japan
        {\tt\small}}%
\thanks{$^{2}$Yoshihiro Kuroda is with the Faculty of Engineering, Information and Systems, University of Tsukuba, Tsukuba, Japan
        {\tt\small}}%
\thanks{$^{3}$Shunsuke Yoshimoto is with the Department of Precision Engineering, School of Engineering, University of Tokyo, Tokyo, Japan
        {\tt\small}}%
}

\maketitle
\thispagestyle{fancy}
\begin{abstract}
\par In recent years, thermal display has been studied intensively in order to represent a more realistic tactile quality of the object. Since human feels the temperature of the air without touching other objects, it is necessary to present thermal sensation in a non-contact manner. Studies on non-contact heat display have been explored; however, few studies have reported on a device that can display cold in a non-contact manner. In this study, we propose a non-contact cold thermal display using a low-temperature heat source---vortex tube, which can generate ultra-low air temperature when supplied with compressed air. We developed a cooling model that relates the flow velocity of cold air with the absorbed heat from skin; we implemented a prototype system that can control the flow velocity of the generated air; and we conducted an experiment to examine the cold sensation that the system can present. Our results revealed that various cold sensations can be generated so that the faster the flow velocity, the colder a user would feel.
\end{abstract}

\section{Introduction}
\par When we travel to places that we have never visited, we recognize the situation by seeing, hearing, smelling, or feeling the surrounding air. For example, if we see a white world, hear the sound of wind and feel cold, we will realize that we are in snow. In recent years, with the spread of head mounted display (HMD) and stereophonic sound, visual and auditory information can be generated; consequently, expectations to present skin sensation are increasing for the purpose of giving users a more realistic and natural virtual reality (VR) experience. Skin sensation combines of tactile sensation, pain sensation and thermal sensation. Thermal sensation is an important factor, particularly when recognizing surrounding situations. Since human feels the temperature of the air without touching other objects, it is necessary to present thermal sensation in a non-contact manner. Studies displaying heat in a non-contact manner have been investigated; however, few studies have reported on a device that can display cold without contact. 
\par Cold sensation is related to the amount and speed of heat dissipating from the body; thus, the main factors that affect cold sensation are temperature and flow velocity of the environment air, which determine heat dissipation \cite{coldsensation}. In this study, we propose a system that presents non-contact cold sensation by changing the air temperature and controlling the air flow velocity around the user. To generate cold air, we used a vortex tube, a low-temperature heat source that can generate ultra-low air temperature (e.g. -20~$^\text{o}C$) when supplied with compressed air \cite{vortextube}. We aim to present various cold sensations by controlling the flow velocity of the generated air. 
\par In this paper, we propose a cooling model that relates the flow velocity of cold air with the absorbed heat from skin and we describe our prototype system. In addition, we conducted an experiment to evaluate the cold sensation that the prototype system can present, and the results are reported in this paper.

\section{RELATED STUDIES}
\par In the conventional method, hot and cold sensations are presented in a contact manner to simulate the thermal sensation when touching an object. Many studies present hot and cold sensations using Peltier devices, which can transfer heat from one side of the device to the other depending on the direction of the current and which can be used as a temperature controller that either heats or cools \cite{contact1}\cite{contact2}\cite{contact3}\cite{contact4}\cite{contact7}\cite{contact8}. Sakaguchi et al. \cite{contact5}\cite{contact6} suggest an approach to represent hot and cold sensation by using water for a heat medium. By switching hot/cold water flow using a solenoid valve, one can rapidly change the temperature of the presentation. 
\par However, human feels the temperature of the air without touching other objects; thus, it is necessary to present thermal sensation in a non-contact manner. Studies on displaying heat in a non-contact manner have been investigated. For instance, Hasegawa et al. \cite{noncontact0}\cite{noncontact1} developed a heat display in a non-contact method by using infrared rays to generate heat. Hokayama et al. \cite{noncontact2} suggested a method of presenting heat sensation by changing the humidity of air and the results showed that highly humid air was felt as warmer than normal air. For displaying cold sensations without contact, Nakajima et al. \cite{noncontactcold} suggest to generate and control air flow driven by an ultrasound-phased array and transport cold air to a localized spot on the user's skin. However, it is hard to give user an immersive experience in a cold situation since there is only one spot on the user's skin. In conclusion, it can not present stable and multi-level cold sensations.

\section{COOLING MODEL}
\par We developed a cooling model that relates the flow velocity of cold air and the heat absorbed from skin. Note that the heat of the cold air would be quickly transferred to the skin, because the specific heat capacity of air is rather small compared to that of the skin. We assumed that when cold air comes into contact with the skin, the air temperature rises and shortly becomes the same as the skin temperature.

\begin{figure}[htbp]
	\centering
	\includegraphics[scale=0.24]{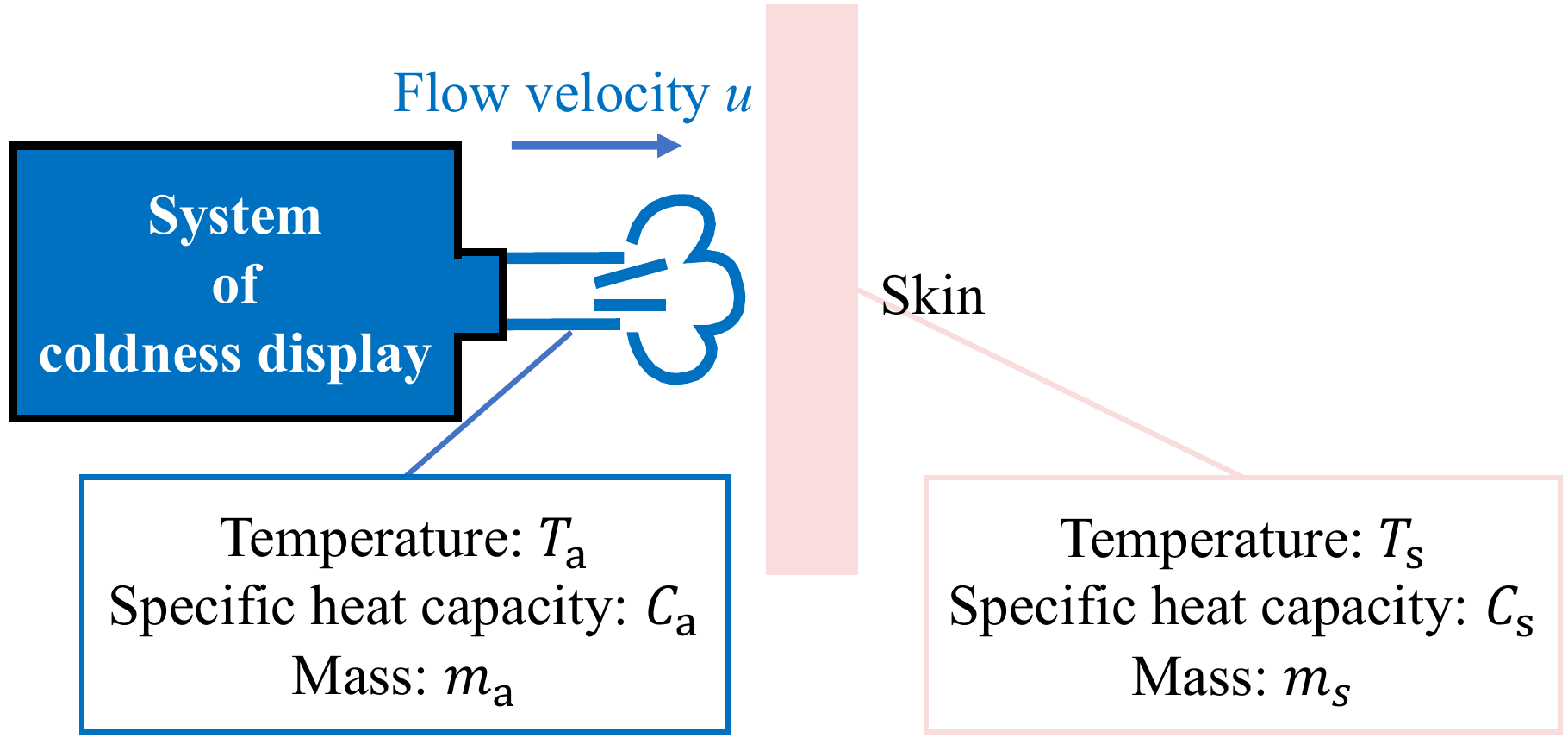}
	\caption{Cooling model}
	\label{fig:Cooling model}
\end{figure}
\par As shown in Fig.~\ref{fig:Cooling model}, our model has the following assumptions:
\begin{itemize}
	\item The specific heat capacity $C_{\rm a}$ and the density $\rho_{\rm a}$ of the cold air are constant. The initial temperature of the cold air is $T_{\rm a}$.
	\item The specific heat capacity $C_{\rm s}$ and the density $\rho_{\rm s}$ of the skin are constant. The initial temperature of the skin is $T_{\rm s}$.
	\item The area $A_{\rm s}$ of the skin in contact with cold air is constant and the thickness of skin is $h_{\rm s}$.
	\item When the cold air reaching the same temperature as the skin $T_{\rm f}$, heat $Q$ is removed from the skin.
\end{itemize}

\par Assuming that the cross-sectional area of the cold air outlet is $A$, the flow rate $K$ is expressed by Eq.~(\ref{eq：体積流量と流速の関係}),
\begin{equation}\label{eq：体積流量と流速の関係}
K=uA
\end{equation}
where $u$ is the speed of cold air flowing in the normal direction of the skin. The volume $V_{\rm a}$ of the output cold air for duration time $t$ is expressed as
\begin{equation}\label{eq：t秒間出力される冷気の体積}
V_{\rm a}=Kt=uAt.
\end{equation}
Therefore, the mass $m_{\rm a}$ of the output cold air for duration time $t$ is expressed as
\begin{equation}\label{eq：t秒間出力される冷気の質量}
m_{\rm a}=\rho_{\rm a}V_{\rm a}=\rho_{\rm a}uAt.
\end{equation}
Consequently, the amount of heat $Q$ deprived from the skin for duration time $t$ is expressed by Eq.~(\ref{eq：皮膚から奪われる熱量}).
\begin{equation}\label{eq：皮膚から奪われる熱量}
Q=C_{\rm a}m_{\rm a}(T_{\rm f}-T_{\rm a})=C_{\rm a}\rho_{\rm a}uAt(T_{\rm f}-T_{\rm a})
\end{equation}
The volume $V_{\rm s}$ of the skin in contact with cold air is expressed as
\begin{equation}\label{eq：冷気と接触する皮膚の体積}
V_{\rm s}=A_{\rm s}h_{\rm s}.
\end{equation}
Therefore, the mass $m_{\rm s}$ of the skin in contact with cold air is expressed as
\begin{equation}\label{eq：冷気と接触する皮膚の質量}
m_{\rm s}=\rho_{\rm s}V_{\rm s}=\rho_{\rm s}A_{\rm s}h_{\rm s}.
\end{equation}
The temperature change $\Delta T_{\rm s}$ of the skin in contact with cold for duration time $t$ can be expressed by Eq.~(\ref{eq：皮膚の温度変化1}).
\begin{equation}\label{eq：皮膚の温度変化1}
\Delta T_{\rm s}=T_{\rm s}-T_{\rm f}=\frac{Q}{C_{\rm s}m_{\rm s}}
\end{equation}
Eq.~(\ref{eq：皮膚の温度変化2-1}) is derived from Eq.~(\ref{eq：皮膚から奪われる熱量}) and Eq.~(\ref{eq：皮膚の温度変化1}).
\begin{equation}\label{eq：皮膚の温度変化2-1}
T_{\rm s}-T_{\rm f}=\frac{C_{\rm a}\rho_{\rm a}uAt(T_{\rm f}-T_{\rm a})}{C_{\rm s}\rho_{\rm s}A_{\rm s}h_{\rm s}}   
\end{equation}
We define a coefficient $k$ in Eq.~(\ref{eq：k}).
\begin{equation}\label{eq：k}
    k=\frac{C_{\rm a}\rho_{\rm a}A}{C_{\rm s}\rho_{\rm s}A_{\rm s}h_{\rm s}}
\end{equation}
Eq.~(\ref{eq：皮膚の温度変化2-1}) can be expressed as
\begin{equation}\label{eq：皮膚の温度変化2}
T_{\rm s}-T_{\rm f}=ku(T_{\rm f}-T_{\rm a})t.
\end{equation}
From Eq.~(\ref{eq：皮膚の温度変化2}), $T_{\rm f}$ can be calculated as \begin{equation}\label{eq：皮膚の温度変化3}
T_{\rm f}=\frac{T_{\rm s}+kuT_{\rm a}t}{kut+1}
\end{equation}
Therefore, $\Delta T_{\rm s}$ can be calculated as
\begin{align}\label{eq：皮膚の温度変化}
\Delta T_{\rm s}=T_{\rm s}-T_{\rm f}=\frac{ku(T_{\rm s}-T_{\rm a})t}{kut+1}
\end{align}
\par The temperature change of the skin after duration time $t$ is shown to be related to the magnitude of the flow velocity $u$ and the starting temperature difference $(T_{\rm s}-T_{\rm a})$ between the skin and the cold air.

\section{NON-CONTACT COLD THERMAL DISPLAY}
\par We implemented a prototype of non-contact cold thermal display by changing the flow rate of cold air, since the cold sensation is related to the flow velocity and to the temperature of the air. For constructing a compact display, we used a vortex tube, a small and simple device with no moving parts, to generator cold air. In addition, compared with Peltier heat pump, which needs several time to reach the set temperature, the vortex tube has a good time response as it can generate ultra-low air temperature (e.g. -20 $^\text{o}$C)  when supplied with compressed air. By controlling the flow velocity of the generated air, we aimed to present various cold sensations.
\subsection{Low-temperature heat source: vortex tube}
\begin{figure}[htbp]
	\centering
	\includegraphics[scale=0.24]{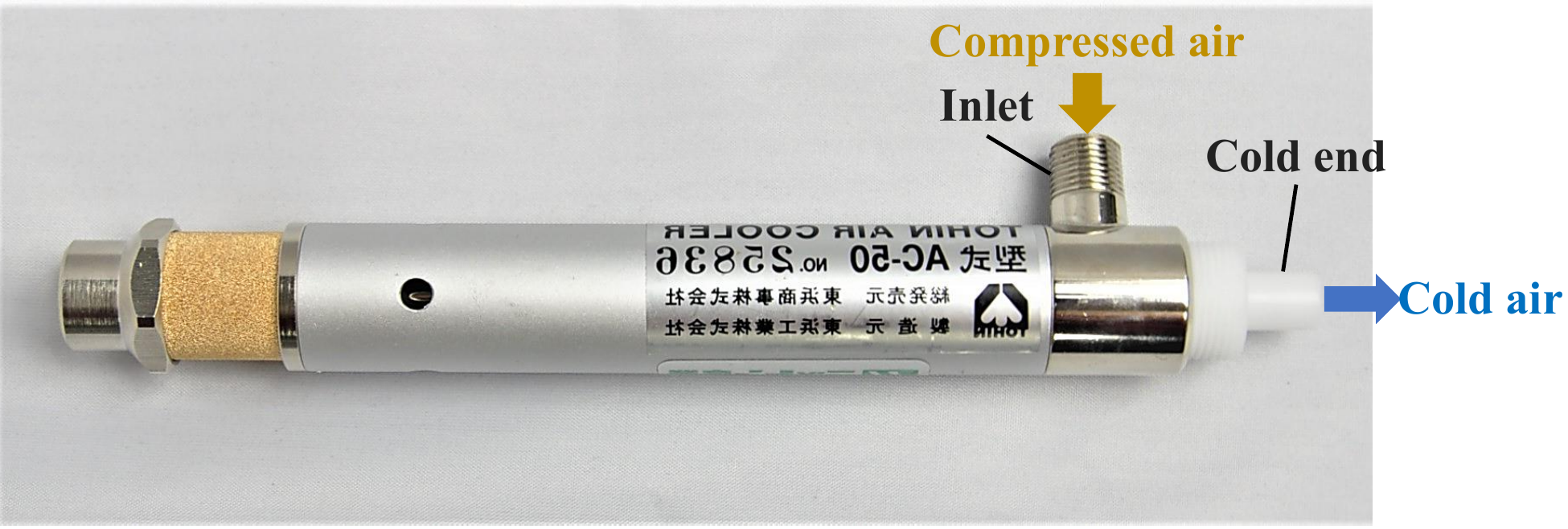}
	\caption{Vortex tube}
	\label{fig:Vortex tube}
\end{figure}
\par The temperature of the generated air is related to the pressure and the temperature of the supplied air. The higher the pressure of the supplied air is, the lower the temperature of the generated cold air is; the lower the temperature of the supplied air is, the lower the temperature of the generated cold air is.
\subsection{Overview}
\par
\begin{figure}[htbp]
	\centering
	\includegraphics[scale=0.24]{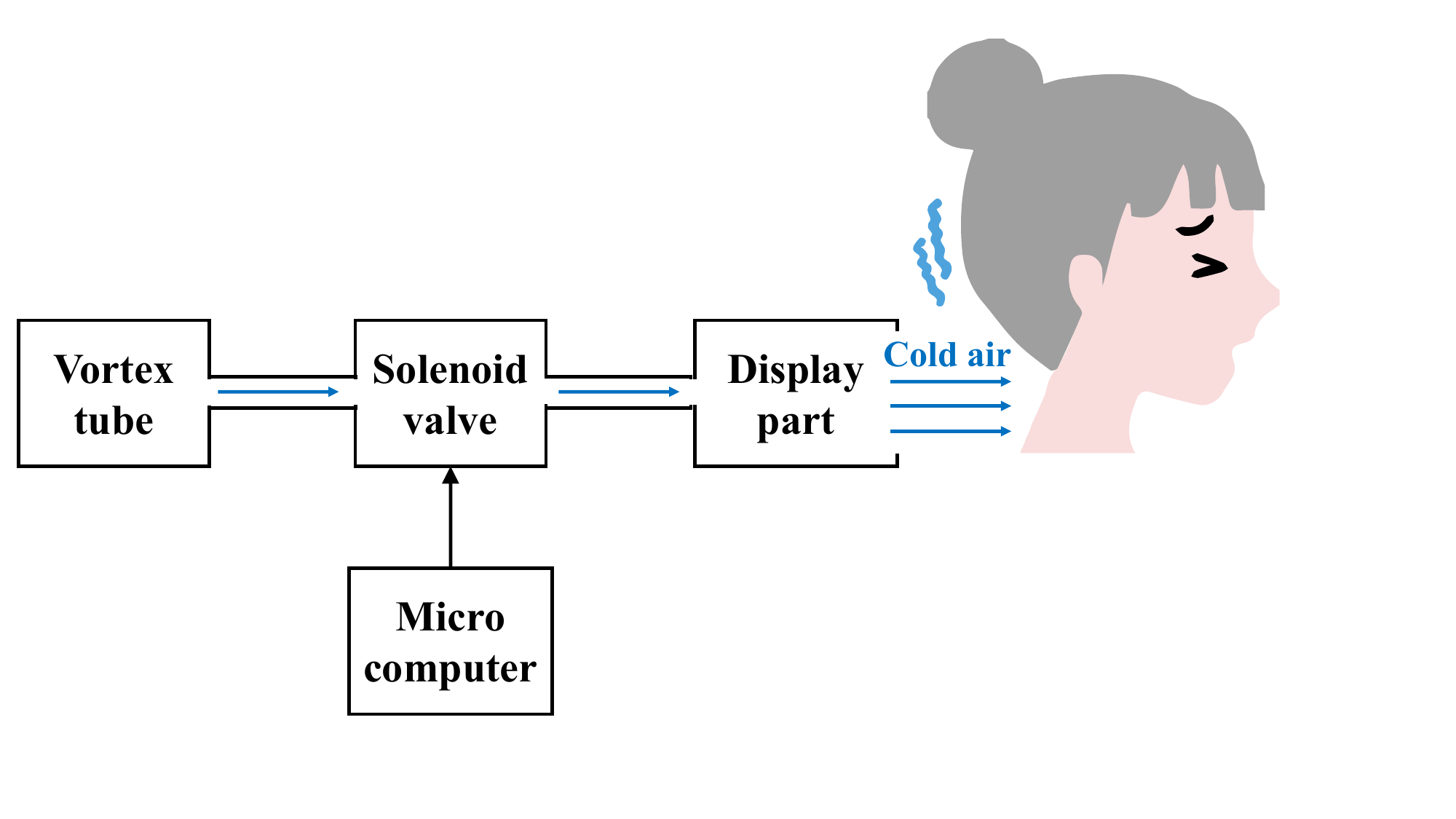}
	\caption{A schema of prototype system}
	\label{fig:A schema of prototype system}
\end{figure}

 Fig.~\ref{fig:A schema of prototype system} shows a schema of the prototype system. The vortex tube generates cold air. Then, using a micro-computer, the solenoid valve changes the amount of passing generated cold air. Finally, the cold air is displayed to the user through a heat-insulated tube. 
\subsection{System configuration}
\par
\begin{figure}[htbp]
	\centering
	\includegraphics[scale=0.26]{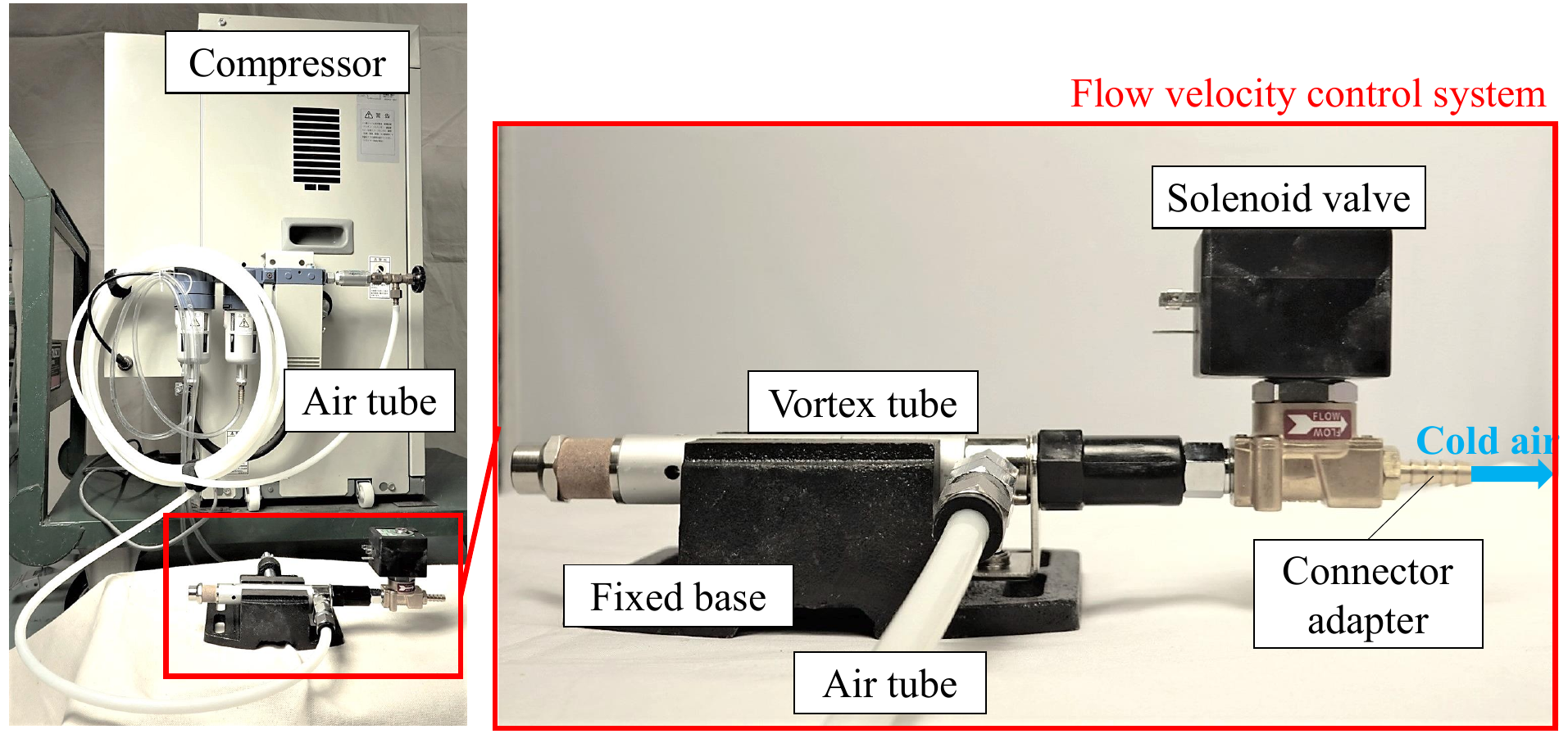}
	\caption{Prototype system}
	\label{fig:Prototype system}
\end{figure}

 Figure \ref{fig:Prototype system} shows the developed prototype system. Vortex tube by Tohin/AC-50 was used as the cold air generator. It is possible to adjust mechanically the cold air ratio of the vortex tube, which is the relative volume of output cold air compared to the input compressed air. When the cold air ratio is high, the amount of output cold air is large, while the temperature difference from the input compressed air is small. In this system, the cold air ratio was set to 50~\%. Compressor by Hitachi/POD-0.75LES was used to supply compressed air for vortex tube. The pressure provided by the compressor is maintained within a certain range (0.6 to 0.8~MPa).
Solenoid valve by Asco/Positive-flow-202 was used to control flow velocity of the generated cold air. 
Originally, the solenoid valve can control the flow rate of the air within the range of 0 to 100~\% by the applied voltage. 
Since the cross-sectional area of the outlet of the solenoid valve is constant, the flow velocity can be controlled by changing the flow rate of the air. 
\par We change the output of the solenoid valve by pulse width modulation (PWM). The control input is the voltage based on duty ratio. The frequency of the PWM was set to 300~Hz, which was the optimal frequency of the valve. 

 \section{EXPERIMENT-1 TEMPERATURE DROP OF THE TEMPERATURE SENSOR}
\par In the first experiment, we investigated the temperature drop that accompanies the change of flow velocity of cold air, using the prototype system.
\subsection{Experimental conditions and settings}
\par
\begin{figure}[htbp]
	\centering
	\includegraphics[scale=0.25]{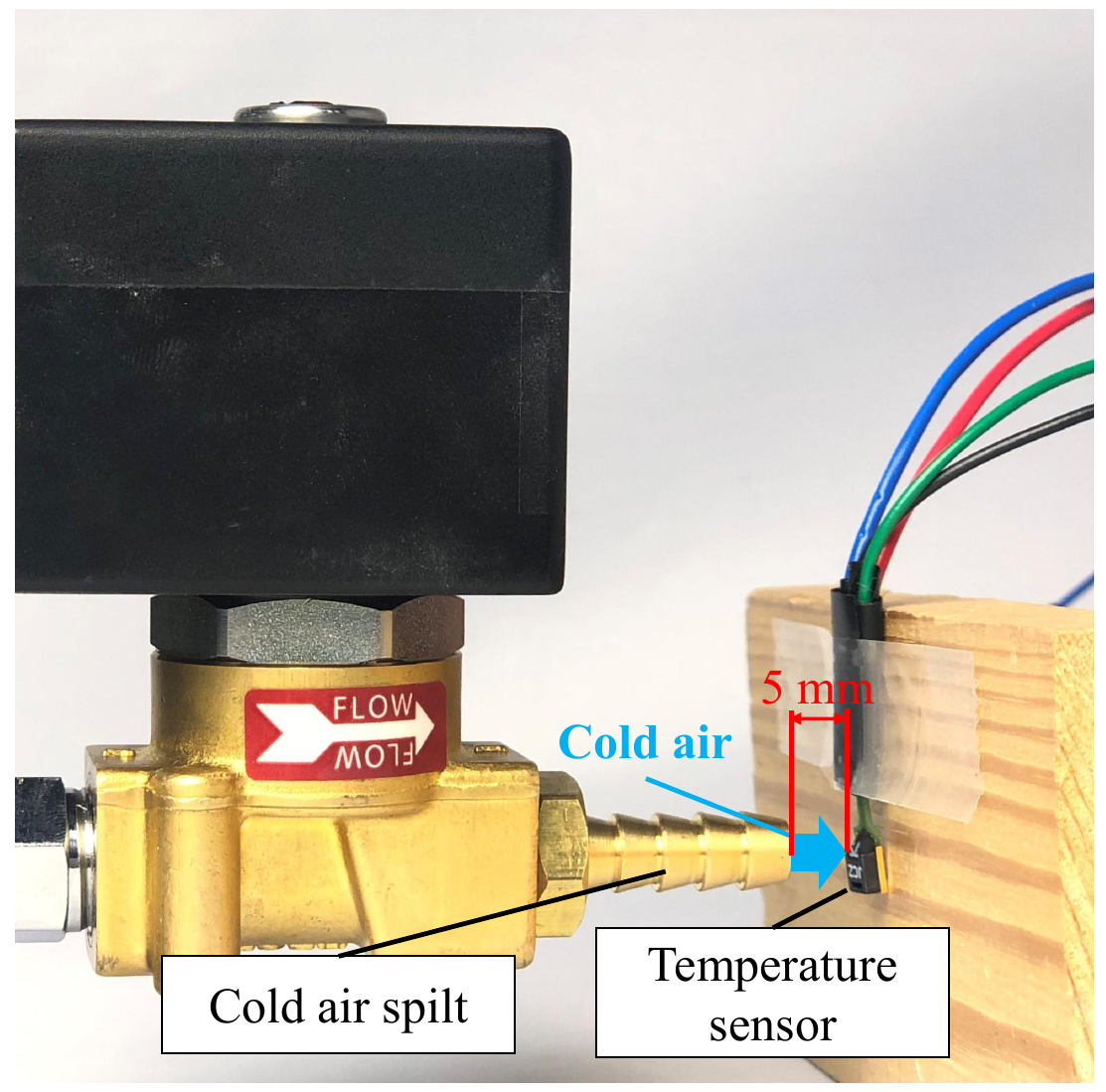}
	\caption{Experiment-1: Conditions and settings}
	\label{fig:Experimental conditions and settings1}
\end{figure}

 As shown in Figure~\ref{fig:Experimental conditions and settings1}, a temperature sensor SHT75 by Sensirion was placed 5~mm from the cold air spilt. Measurements were performed at eight flow velocities (0.0, 0.5, 1.0, 1.5, 2.0, 2.5, 3.0, 3.5~m/s). Due to the distance between the cold air spilt and the temperature sensor, the measurement value was the temperature of the mix of the dispersed cold air and the environment air. During the experiment, the room temperature was 22~$^\text{o}$C. The temperature of the generated cold air (measured by the temperature sensor SHT75 by Sensirion) was -16~$^\text{o}$C.
\subsection{Results}
\par
\begin{figure}[htbp]
	\centering
	\includegraphics[scale=0.27]{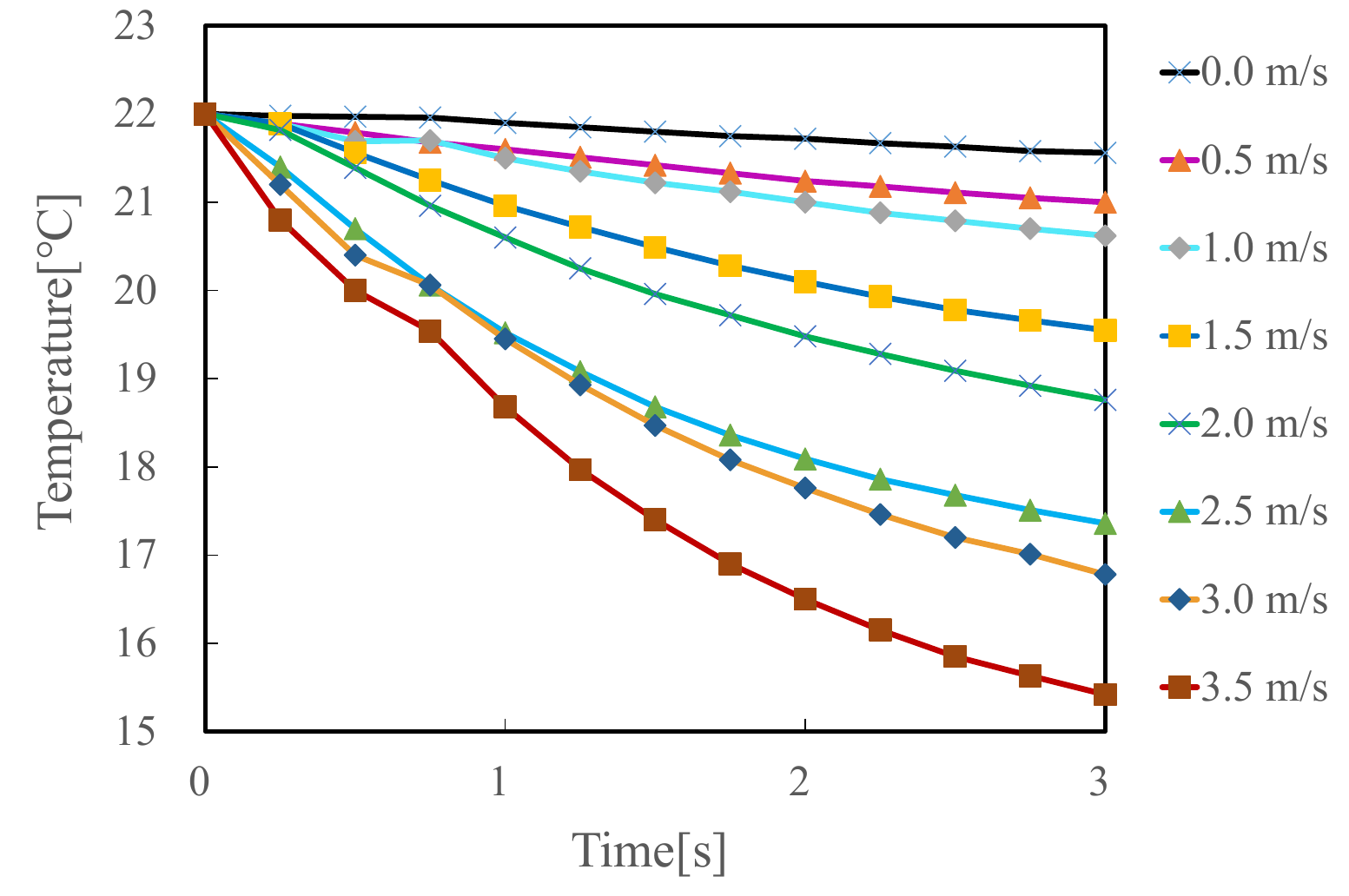}
	\caption{Result of Experiment-1}
	\label{fig:Result of Experiment-1}
\end{figure}

 Figure~\ref{fig:Result of Experiment-1} shows all measured transitions of temperature in 3~s at each flow velocity. The result showed that the faster the flow velocity of the air, the faster the temperature drops.

\section{EXPERIMENT-2 TEMPERATURE CHANGE OF THE SILICON SHEET}
\par In the second experiment, in order to verify the developed cooling model, we investigated the temperature change that accompanies the change of the flow velocity of the cold air using a phantom. 
\subsection{Experimental conditions and settings}
\par
\begin{figure}[htbp]
	\centering
	\includegraphics[scale=0.26]{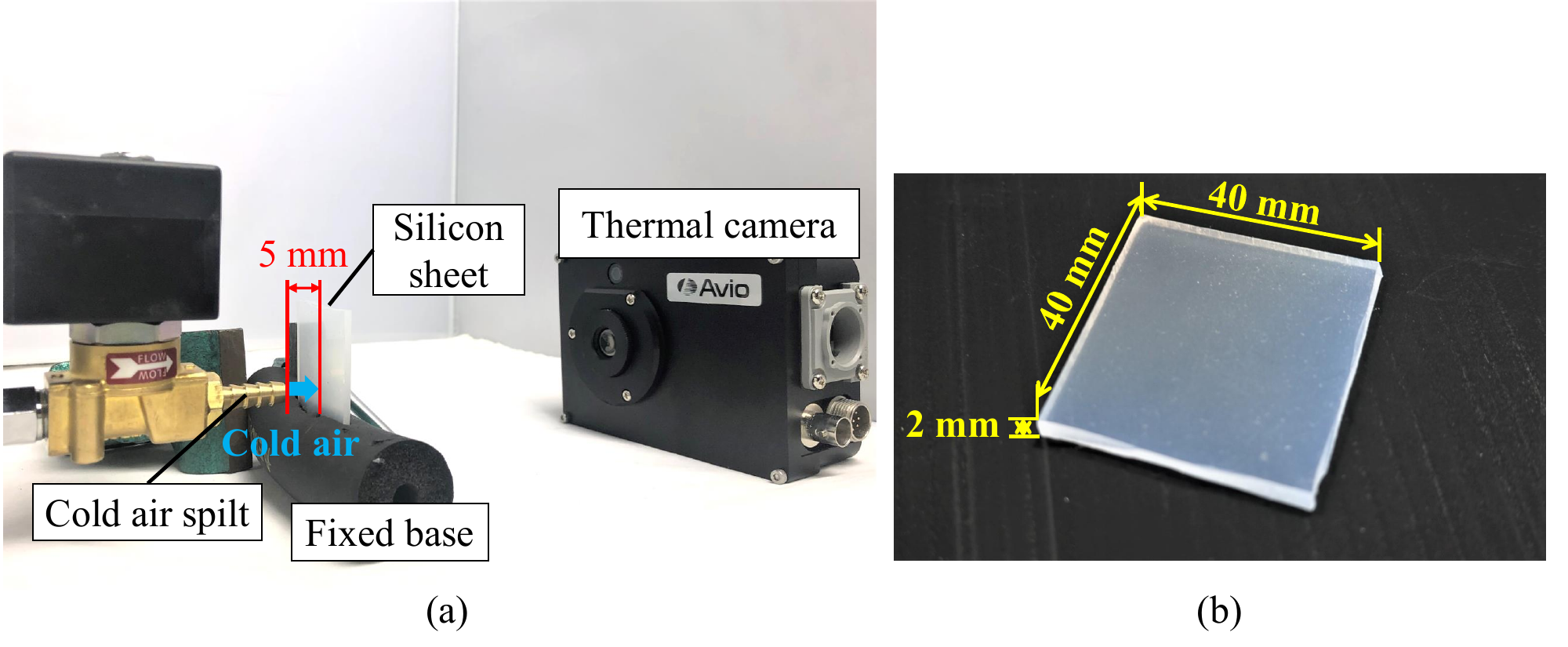}
	\caption{Experiment-1: (a)Conditions and settings (b)Silicon sheet}
	\label{fig:Experimental conditions and settings 1}
\end{figure}

 We used a silicon sheet as a phantom (instead of the skin). Since the heat absorbed by the cold air from the surrounding air (room temperature) increases with the distance between the cold air spilt and the skin, we chose to present cold sensations in a short distance to have an effective cooling effect. As shown in Fig.~\ref{fig:Experimental conditions and settings 1}(a), the silicon sheet was placed 5~mm from the cold air spilt. 
As shown in Fig.~\ref{fig:Experimental conditions and settings 1}(b), the area $A_{\rm s}$ of the silicon sheet was 40~mm$\times$40~mm, or=1600~mm$^{2}$, and the thickness $h_{\rm s}$ of the silicon sheet was 2~mm. The values of specific heat capacity $C_{\rm s}$ and density $\rho_{\rm s}$ of silicon were set to 1.6~kJ/kg $\cdot$ K and 970~kg/m$^{3}$, respectively, which typical for the material. 
The temperature distribution of the silicon sheet was repeatedly measured using a thermographic camera (S30-series by InfReC) with the flow velocity $u$ of cold air controlled to 1, 2 and 3~m/s. 
During the experiment, the room temperature was 22~$^\text{o}$C. The temperature of the generated cold air (measured by the temperature sensor SHT75 by Sensirion) was -16~$^\text{o}$C(257.15~K). 
The specific heat capacity $C_{\rm a}$ of air of 257.15~K is 1.005~kJ/kg $\cdot$ K. The density $\rho_{\rm a}$ of air of 257.15~K is 1.37~kg/m$^{3}$. 
\par Since the cross-sectional area $A$ of the cold air spilt is 19.64~mm$^{2}$, coefficient $k$ can be calculated from Eq.~(\ref{eq：k}) as Eq.~(\ref{eq：Siliconのk}).
\begin{align}\label{eq：Siliconのk}
k&=\frac{C_{\rm a}\rho_{\rm a}A}{C_{\rm s}\rho_{\rm s}A_{\rm s}h_{\rm s}}\notag\\
&=\frac{1.005 \times 1.37 \times 19.64 \times 10^{-6}}{1.6 \times 970 \times 1600 \times 10^{-6}\times 2 \times 10^{-3}}\notag\\
&=0.005
\end{align}
The theoretical value of the temperature change $\Delta T_{\rm s}$ of the silicon sheet after duration time $t$ can be calculated as Eq.~(\ref{eq：Siliconの温度変化}).
\begin{equation}\label{eq：Siliconの温度変化}
\Delta T_{\rm s}=\frac{0.005u(T_{\rm s}-257.15)t}{0.005ut+1}
\end{equation}
\subsection{Results}
\par
\begin{figure}[htbp]
	\centering
	\includegraphics[scale=0.27]{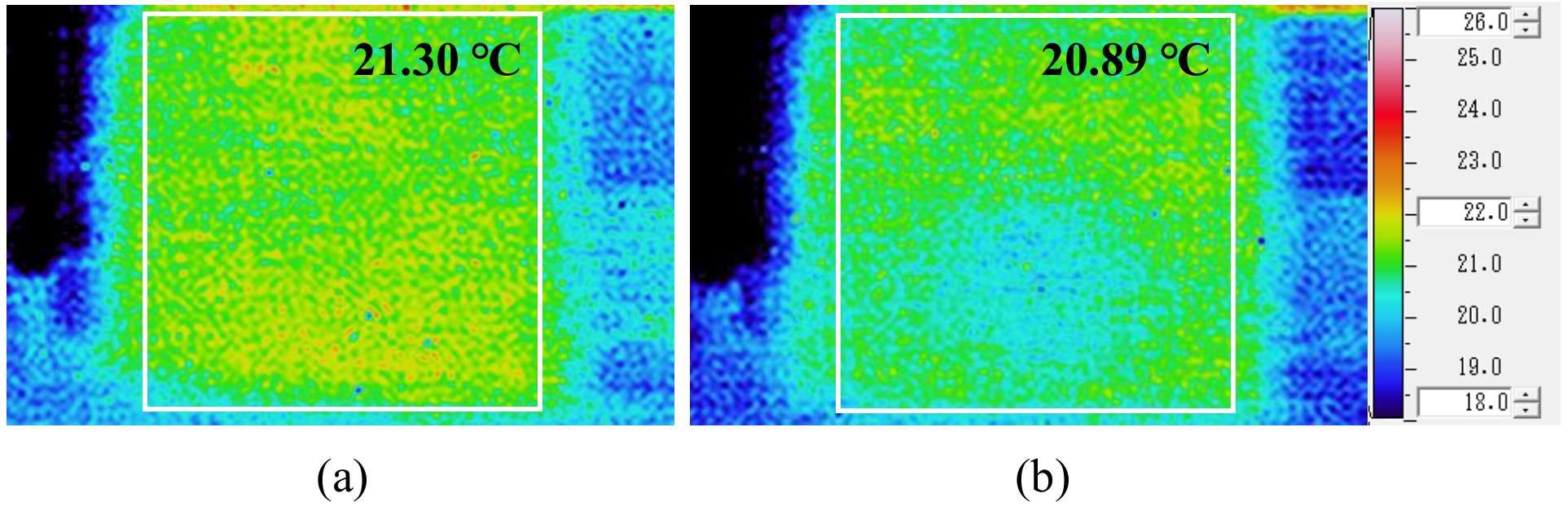}
	\caption{Flow velocity = 1.0~m/s: (a)Starting temperature distribution (b)The temperature distribution after 3~s}
	\label{fig:Result1-1}
\end{figure}

\begin{figure}[htbp]
	\centering
	\includegraphics[scale=0.27]{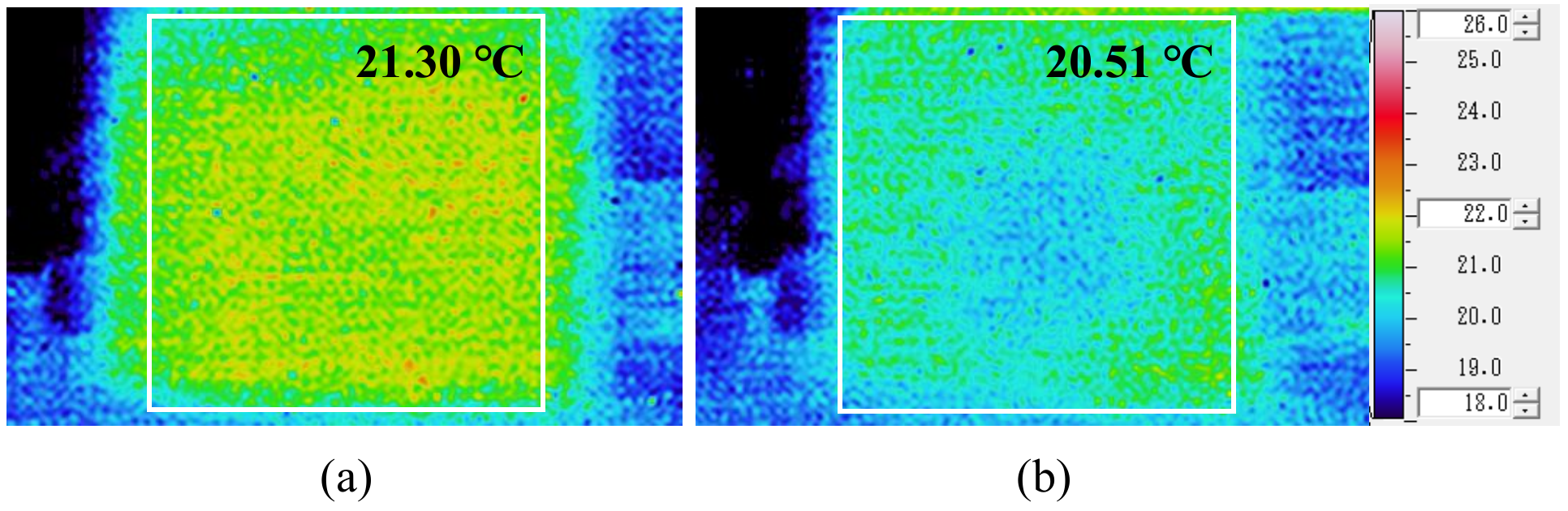}
	\caption{Flow velocity = 2.0~m/s: (a)Starting temperature distribution (b)The temperature distribution after 3~s}
	\label{fig:Result1-2}
\end{figure}

\begin{figure}[htbp]
	\centering
	\includegraphics[scale=0.27]{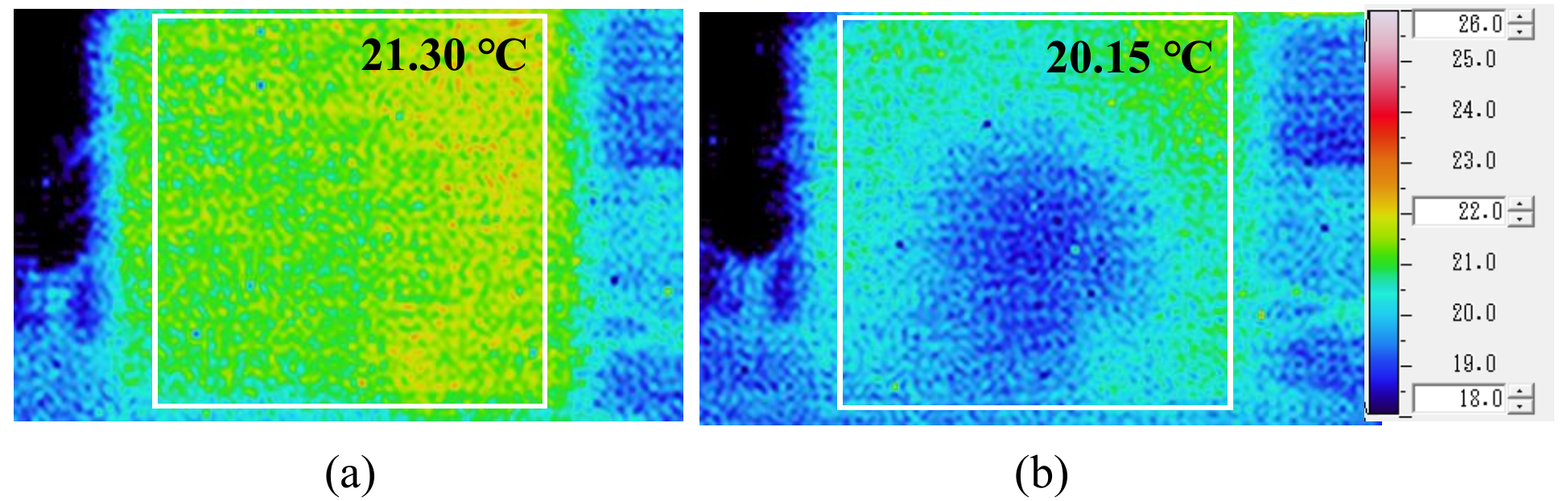}
	\caption{Flow velocity = 3.0~m/s: (a)Starting temperature distribution (b)The temperature distribution after 3~s}
	\label{fig:Result1-3}
\end{figure}

 The results are shown below. The temperature distribution of the silicon sheet is shown in the part surrounded by the white line in Fig.~\ref{fig:Result1-1} to Fig.~\ref{fig:Result1-3}. The software "InfReC Analyzer NS 9500 Lite" was used to read the average spatial temperature of the silicon sheet.
\par As shown in Fig.~\ref{fig:Result1-1}, when the flow velocity was 1.0~m/s and the starting average spatial temperature was 21.30~$^\text{o}$C(294.45~K), the average spatial temperature of the silicon sheet after 3~s was 20.89~$^\text{o}$C(294.04~K). The temperature drop was 0.41~K.
\par As shown in Fig.~\ref{fig:Result1-2}, when the flow velocity was 2.0~m/s and the starting average spatial temperature was 21.30~$^\text{o}$C(294.45~K), the average spatial temperature of the silicon sheet after 3~s was 20.51~$^\text{o}$C(293.66~K). The temperature drop was 0.79~K.
\par As shown in Fig.~\ref{fig:Result1-3}, when the flow velocity was 3.0~m/s and the starting average spatial temperature was 21.30~$^\text{o}$C(294.45~K), the average spatial temperature of the silicon sheet after 3~s was 20.15~$^\text{o}$C(293.30~K). The temperature drop was 1.15~K.

\subsection{Discussion}
\par
\begin{figure}[htbp]
	\centering
	\includegraphics[scale=0.26]{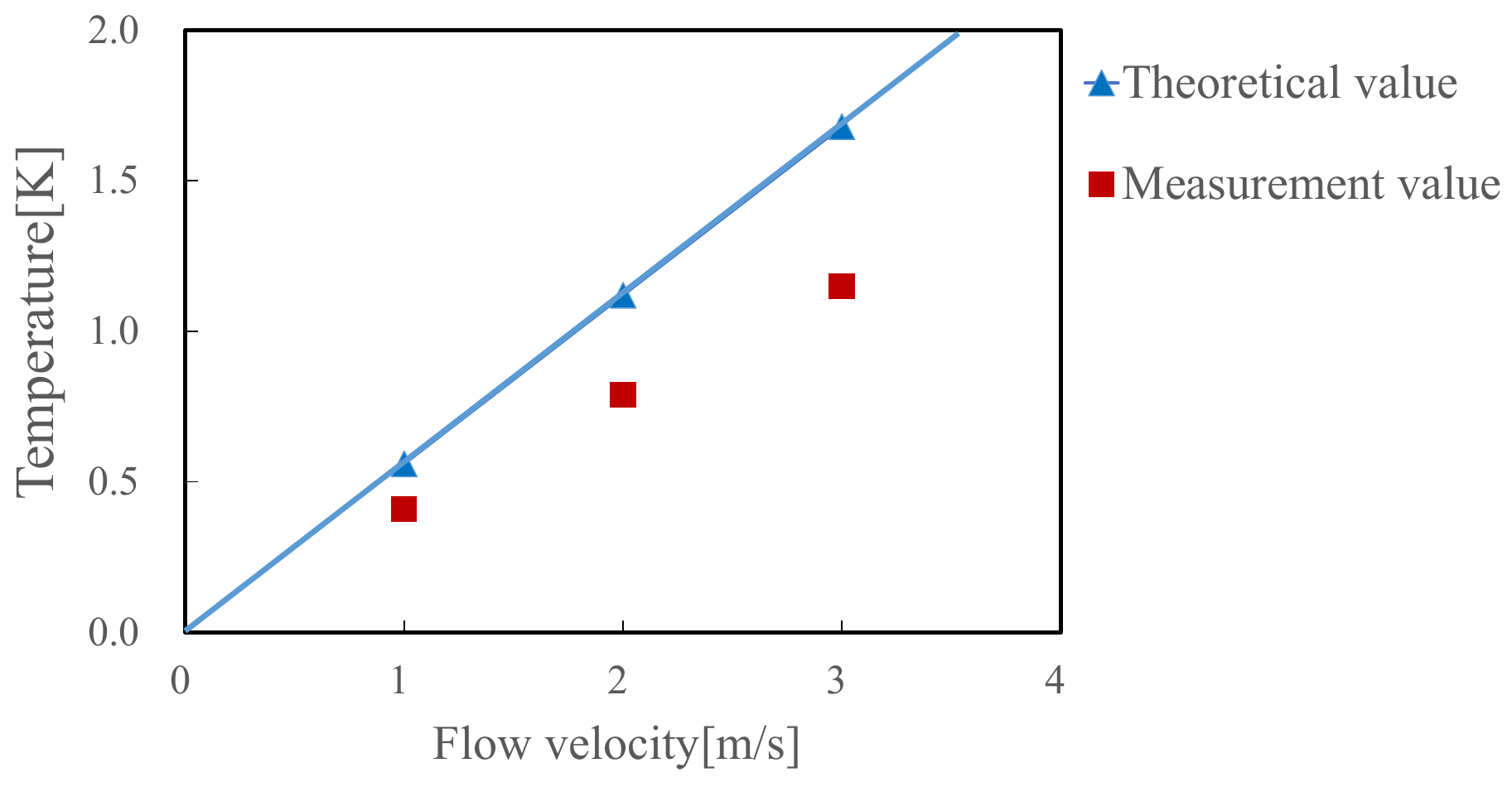}
	\caption{The comparison of the theoretical value and the measurement value}
	\label{fig:The theoretical value and the measurement value}
\end{figure}

 Our results shows that the measured temperature change increases as the flow velocity of the cold air increases. Using Eq.~(\ref{eq：Siliconの温度変化}), we calculated the theoretical value of the temperature of the silicon sheet. The comparing of the theoretical value and the measurement value is shown in Fig.~\ref{fig:The theoretical value and the measurement value}. We assume that the exchange of heat between the cold air and the surrounding air increases with the flow velocity of the cold air, thus the error between the theoretical value and the measured value increases. In addition, the changes of the air velocity at the skin (or the phantom) and the contact area on the skin (or the phantom) could result in fractional changes from the model. In the future, we will consider those effects to compensate our cooling model.
\par The values of specific heat capacity $C_{\rm s}$ and density $\rho_{\rm s}$ of our skin are 10.514~kJ/kg $\cdot$ K and 1200~kg/m$^{3}$, respectively. Since the cold receptors are located directly under the epidermis at a depth of 0.15 to 0.17~mm, we assume the thickness $h_{\rm s}$ of skin in contact with cold air to be 0.20~mm. The temperature of our skin is almost 33~$^\text{o}$C(306.15~K). We assume that the area $A_{\rm s}$ of the skin is 40~mm$\times$40~mm, or=1600~mm$^{2}$, the temperature $T_{\rm a}$ of the cold air is 257.15~K (the same with the silicon sheet). From Eq.~(\ref{eq：k}), coefficient $k$ can be calculated as Eq.~(\ref{eq：皮膚のk}).
\begin{align}\label{eq：皮膚のk}
k&=\frac{C_{\rm a}\rho_{\rm a}A}{C_{\rm s}\rho_{\rm s}A_{\rm s}h_{\rm s}}\notag\\
&=\frac{1.005 \times 1.37 \times 19.64 \times 10^{-6}}{10.514 \times 1200 \times 1600 \times 10^{-6}\times 0.2 \times 10^{-3}}\notag\\
&=0.007
\end{align}
The theoretical value of the temperature change $\Delta T_{\rm s}$ of the skin after duration time $t$ can be calculated as Eq.~(\ref{eq：皮膚の温度変化6}).
\begin{align}\label{eq：皮膚の温度変化6}
\Delta T_{\rm s}&=\frac{0.007u(306.15-257.15)t}{0.007ut+1}\notag\\
                &=\frac{0.34ut}{0.007ut+1}
\end{align}
Therefore, if the flow velocity $u=3$~m/s, the theoretical value of the skin temperature change will be 2.88~K after 3~s. In the future, we will conduct an experiment to verify the theoretical value of skin temperature change.

\section{EXPERIMENT-3 COLDNESS DISCRIMINATION EXPERIMENT BY CONSTANT METHOD}
\par In the final experiment, we conducted a coldness discrimination experiment by constant method to evaluate the cold sensation that the prototype system can present.
\subsection{Participants}
\par 5 healthy volunteers (aged 21–28 years, 1 woman, 4 men) took part in the experiments. None of them were involved in developing our project or knew the purpose of the experiments. The recruitment of the participants and the experimental procedures were approved by the Ethical Committee of the Graduate School Of Engineering Science (30-20), Osaka University, Japan. All participants gave written informed consent to participate and were unaware of the aim of the study. 
\subsection{Experimental conditions and settings}
\par We follow the procedure of the constant method. The flow velocity of the generated cold air was regarded as stimulus. Based on the result of Experiment-1, we selected one flow velocity (2.0~m/s) as Standard Stimulus (SS), and seven flow velocities (0.5, 1.0, 1.5, 2.0, 2.5, 3.0, 3.5~m/s) as Comparison Stimuli (CS). 
\begin{figure}[htbp]
	\centering
	\includegraphics[scale=0.23]{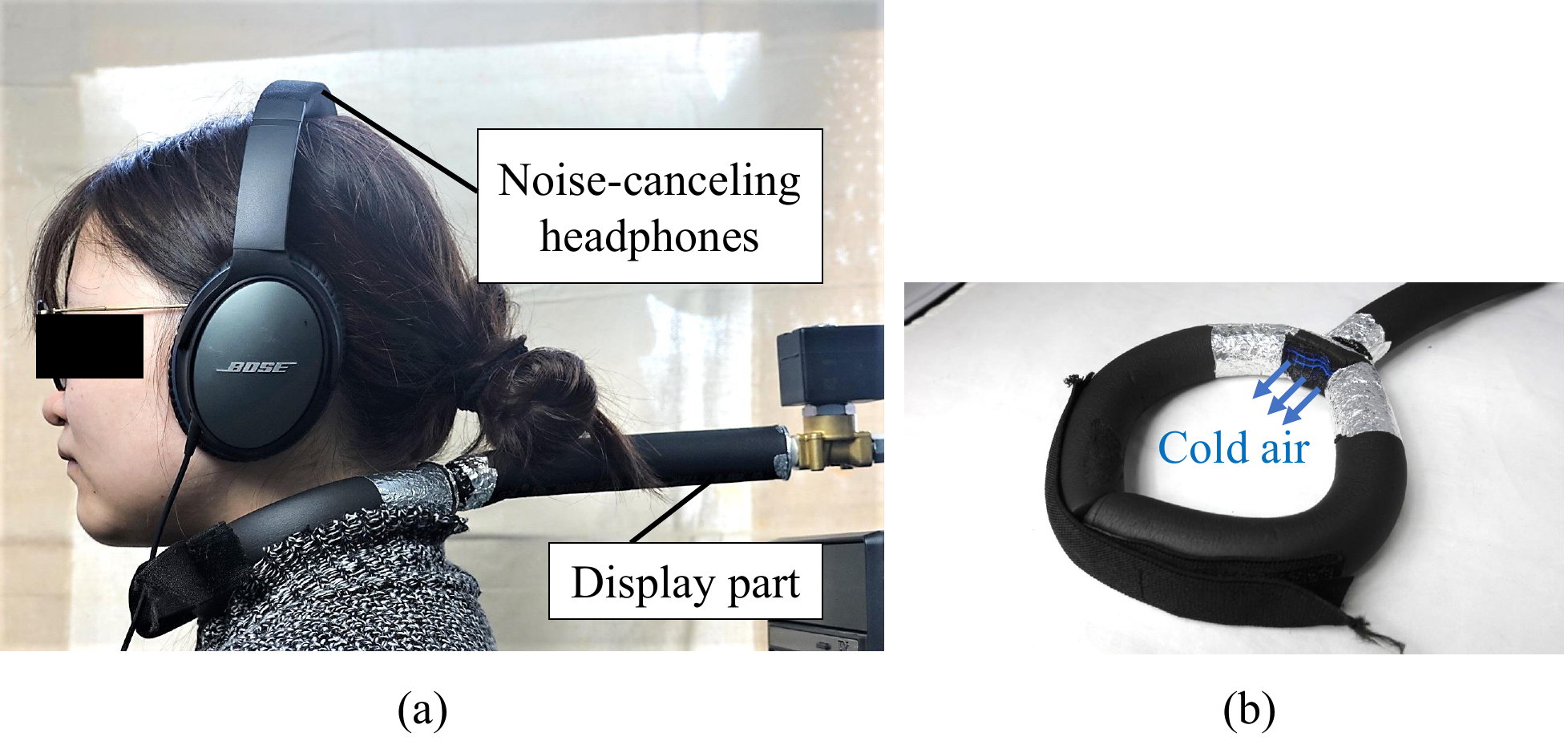}
	\caption{A participant: (a) whole setup and (b) display part}
	\label{fig:Setup of Experiment-2}
\end{figure}
\par During the experiment, participants wore a pair of noise-canceling headphones, and heard pink noise to avoid the sound influence of the compressor, the solenoid valve and the environmental noise. As shown in Figure~\ref{fig:Setup of Experiment-2}, during the presentation of the stimulus, participants wore the display part under a constraint so that the head and body did not move and only looked at the computer screen, which only showed the information of the experiment to let participants focus on the presented cold stimulus. We chose to present cold sensations to the neck in this experiment. Cheeks, neck, and wrists are body parts that are not typically covered by clothes in daily life. In these three parts, the cheek is the most sensitive to cold sensation, the neck is the second and the wrist is the third\cite{thermalsensitivity}. Although the cheek is the most sensitive part, if we display cold sensations on the cheek, we feel that the air would be felt as coming from the front of the face; while it is possible to display cold sensations from all the direction around the neck, so that users could have the feeling of being immersed in an environment.
\par In experiments, each comparison stimulus was randomly presented for ten trials, resulting in 7 stimulus x 10 trials = 70 trials in total. In each trial, the first stimulus was first presented for about 2~s; then, the solenoid value was turned off for about 1~s, to stop presenting the stimulus. Subsequent stimulus was presented for 2~s, and when the presentation was completed, the participants answered which stimulus they felt as colder. We randomly decided which was the comparison stimulus and which was the standard stimulus. In order to avoid unreliability of answers due to fatigue, we decided to provide a break time every 10 trials, and the device could be removed during the break time. During the experiment, the room temperature was 22~$^\text{o}$C.

\subsection{Results}
\par
\begin{figure}[htbp]
	\centering                                                      
	\subfigure{                   
		\begin{minipage}{5.7 cm}
			\centering                                              
			\includegraphics[scale=0.3]{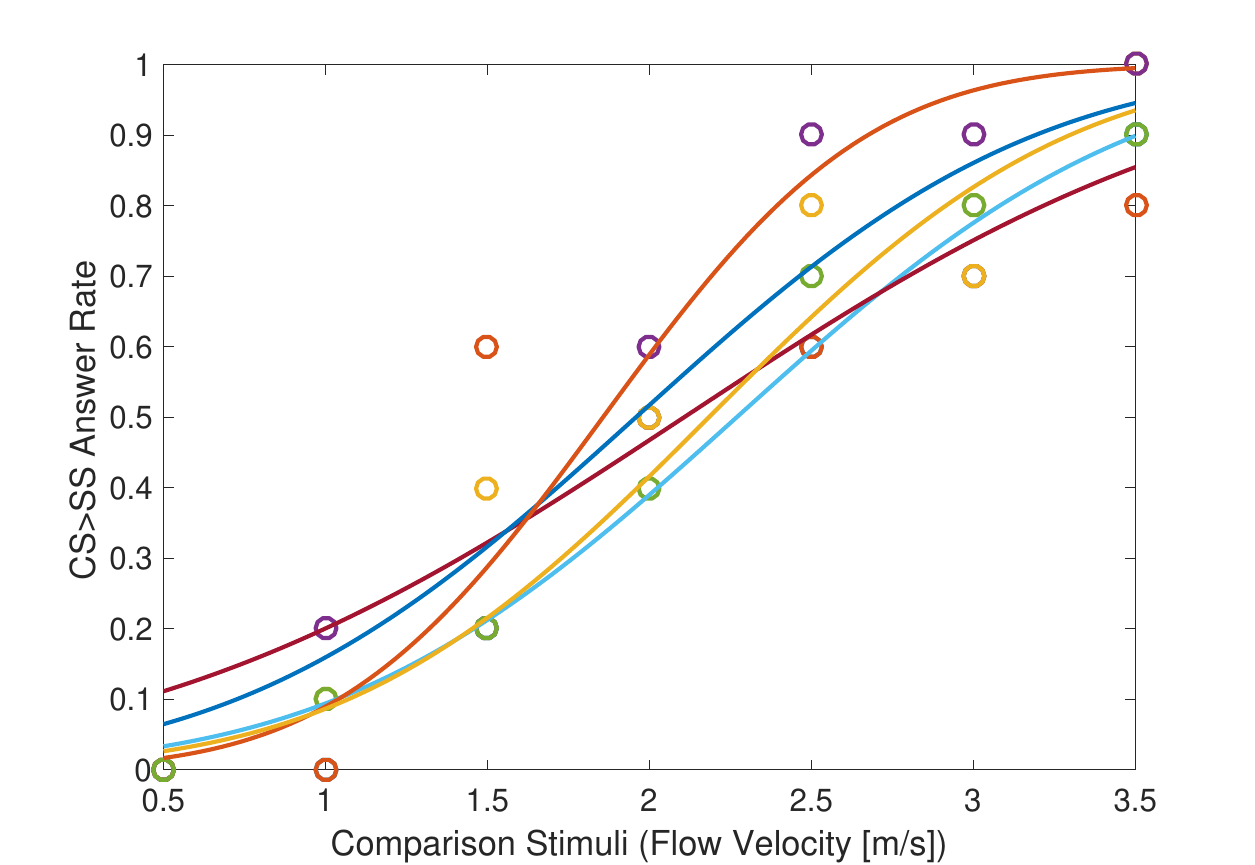}      
		\end{minipage}}
	\subfigure{              
		\begin{minipage}{2.2 cm}
			\centering                       
			\includegraphics[scale=0.3]{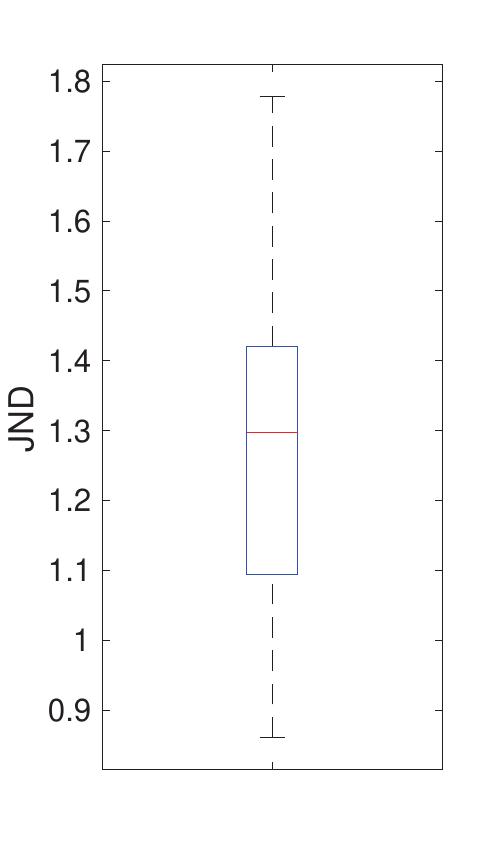}   
		\end{minipage}}
	\caption{Result of Experiment 3} 
	\label{fig:An example of the result}                              
\end{figure}

 To evaluate the discrimination performance, we calculated the probability of a participant answering that the comparison stimulus was colder. Fig.~\ref{fig:An example of the result} shows the result of fitting psychometric curves to the probabilities using a normalized
cumulative distribution function. To quantitatively evaluated the discrimination performance, we calculated the just-noticeable difference(JND) as half the difference between the inverse of the curves at the 0.75 and 0.25 threshold levels. The average and the standard deviation of the JND of the 5 participants were 1.2818~m/s and 0.3304~m/s, respectively. Fig.~\ref{fig:An example of the result} showed the box plot of the JND results.
Therefore, we showed the result that the faster the flow rate is, the colder users will feel; thus, we suggested that multi-level cold sensation can be presented with our prototype system. 

\section{CONCLUSION}
\par In this study, we aimed to represent cold sensation in a non-contact manner. In a cold situation, the thermal sensation is primarily related to the flow velocity and the temperature of the surrounding air. Therefore, we developed a cooling model that relates the flow velocity of cold air with the absorbed heat from skin, and we implemented a prototype system. We used a vortex tube to generate cold air and aimed to present various cold sensation by controlling the flow velocity of the generated air.
\par In the future, we will develop a non-contact cold thermal display by separately controlling the temperature and the flow velocity of the air, to generate more varied cold sensations. 
The new non-contact cold thermal display will provide unique VR experiences such as virtual traveling to Antarctica, where it was almost impossible to travel before. In addition, we will find a high-temperature heat source such as hair dryer to generator hot air. By mixing the cold air and hot air, it is possible to present multi-level thermal sensation.

\end{document}